# DISCUSSION OF: STATISTICAL ANALYSIS OF AN ARCHAEOLOGICAL FIND

By Randall Ingermanson

*Ingermanson Communications, Inc.*

We critique the analysis by A. Feuerverger of an archaeological find that has been alleged by some to be the tomb of Jesus of Nazareth. We show that his analysis rests on six faulty assumptions that have been severely criticized by historians, archaeologists, and scholars in related disciplines. We summarize the results of an alternative computation using Bayes' theorem that estimates a probability of less than 2% that the Talpiot tomb belongs to Jesus of Nazareth.

**1. Introduction.** Andrey Feuerverger notes in his article that assumptions A.1 through A.9 are "not universally accepted." We argue that most historians and archaeologists actually disbelieve his key assumptions. (We agree with Feuerverger that the computational method he proposes can be extremely useful for difficult problems such as the Talpiot tomb.)

Assumption A.7 (the largest driver of his results) is almost universally rejected by scholars in the relevant fields. Several other assumptions are extremely dubious, and *each* of them biases the result toward $H_1$. Since all statistical biases in Feuerverger's RR values accumulate *multiplicatively*, the net effect is an enormous bias toward $H_1$.

In this article, we will look first at the most egregious problem, the "Mariamenou" inscription, which Simcha Jacobovici identified with Mary Magdalene through a long chain of reasoning that has been severely criticized by historians. In less detail, we will examine five other serious problems. By Feuerverger's own account, eliminating two of these statistical biases (the two relating to Mary Magdalene) is sufficient to destroy the statistical significance of $H_1$. But all six statistical biases should be eliminated from the baseline model of the problem.

We describe a series of calculations using Bayes' theorem that show that the probability that the tomb belongs to Jesus of Nazareth is at most about 2%, and may be much less.

**2. The primary problem: The "Mariamenou" inscription.** One of the ossuaries bears an inscription that is usually translated "Mariamenou [who is also called] Mara." Simcha Jacobovici (2007) took this to be a variant of "Mariamne" and interpreted it as a reference to Mary Magdalene. That is, he believed Mary Magdalene went by this name and that very few other women did. Jacobovici based his theory on the work of Dr. Francois Bovon. But Bovon (2007) immediately repudiated this interpretation of his work in a web article. The key point is this statement: "I do not believe that Mariamne is the real name of Mary of Magdalene."

Dr. Richard Bauckham (2007), a renowned expert in first-century Jewish names, has analyzed the "Mariamenou" inscription in detail in a guest blog article. His conclusions are:

(1) Grammatically, "Mariamenou" is the genitive case of the rare form "Mariamenon," a diminutive endearment deriving from the common name "Mariam."
(2) The name is not derived from "Mariamne."
(3) The name is very rare, and no other instance is found in antiquity.
(4) We have no evidence that Mary Magdalene ever went by this name.

One should ask what name Mary Magdalene went by, according to the data we have. Stephen Pfann (2007) has tabulated the references to Mary Magdalene in the various books of the New Testament, the earliest sources that mention her. She is called by the formal name "Mariam" four times and by the shorter, more intimate form "Maria" 10 times. These are the *only* names used in the New Testament to refer to Mary Magdalene.

With these facts at hand, we can answer the following question: Assuming that Mary Magdalene was actually buried in ancient Jerusalem, if one finds the inscription "Mariamenou" in that city, what is the probability that it might refer to Mary Magdalene? The answer is that the inscription is neither more nor less likely to refer to Mary Magdalene than to any other Mary of Jerusalem. (There were roughly 8500 other Marys.) This demolishes Jacobovici's theory, because "Mariamenou" simply can't be identified as "the real name" of Mary Magdalene.

In Feuerverger's article, he assigns an RR value to the Mariamenou inscription that carries an illicit factor of $(1/44)$, due to his belief that the inscription "represents the most appropriate specific appellation for Mary Magdalene from among those known." But it doesn't, and therefore this factor $(1/44)$ should be changed to 1.

This faulty assumption biases the entire calculation very strongly toward $H_1$ and is the primary driver behind the allegedly remarkable results.

**3. Five other significant problems.** In addition to the "Mariamenou" issue, there are a number of other problems in Feuerverger's work that bias the computation toward $H_1$. Each of them contributes a factor smaller than 1. The result of multiplying them all together is an enormous bias toward $H_1$. These problems are as follows:

(1) Assumption A.3 asserts that "the most appropriate rendition of the name of the mother is Marya." Note that "Marya" is the short form of the more formal name "Mariam" and is often spelled "Maria" in English. Assumption A.3 asserts that the mother of Jesus could *not* be listed as "Mariam" on her ossuary. With this assumption, Feuerverger inserts a factor of (13/44) into his RR value for the "Maria" inscription. The problem is that there is no compelling reason to believe A.3. The New Testament data compiled by Stephen Pfann (2007) shows that the mother of Jesus was called "Mariam" 13 times and "Maria" six times. So the data runs counter to Feuerverger's assumption. The mother could be called by either name. Feuerverger's factor of (13/44) is illicit and should be eliminated.

(2) Assumption A.3 likewise asserts that the short form "Yoseh" is the most appropriate rendition of the second brother of Jesus, whose formal name was "Yehosef" like his father. The New Testament refers to this brother once by the short form and once by the long form. A complicating factor here is that any randomly chosen "Yehosef son of Yehosef" would be quite likely to carry an alternative form of the name, so as to distinguish between father and son. Feuerverger inserts a factor of (7/46) into his RR value, which is too small, because it is at the minimum of the range of possible values. The correct value should lie somewhere between (7/46) and 1.

(3) An inscription "Judah son of Jesus" indicates that the Jesus buried in the tomb had a son. Jewish men of the time were very likely to be married and have children. But it is probable that Jesus had no sons. Recall that Jesus had four brothers who assumed positions of influence in the early Jesus movement. If a son also existed, he would likely have joined his uncles in a position of influence and we would have heard of him. Since we have not, we can conclude that the probability that Jesus had a son is *lower* than the probability for a randomly selected man of Jerusalem. Feuerverger's calculation fails to account for this. This inserts a bias into his computation.

(4) If the Talpiot tomb contained the *family* of Jesus of Nazareth, *would we expect Jesus to be in it*? Archaeologist Jodi Magness (2007) argued from a historical perspective that we should not. (But note James Tabor's rebuttal (2007), which argues that the tomb "should not be dismissed." We agree that it should not be dismissed, but it must stand on its merits.) Magness and Tabor at least agree that the data indicates that the body of Jesus went missing within days after the crucifixion. The earliest Jesus movement explained this by asserting that Jesus was resurrected, a claim outside the bounds of scientific investigation. If one looks for a naturalistic explanation, Magness says that much the likeliest one is that Jesus was reburied in a simple trench grave like other poor men of his time. She argues on several grounds that it is implausible that Jesus was buried in a rock-cut tomb like the one at Talpiot. Feuerverger's analysis fails to penalize $H_1$ on account of this issue, thereby introducing another source of statistical bias into his calculations.

(5) Would Mary Magdalene be buried in the family tomb of Jesus? According to Bauckham (2007), the usual practice was that only family members were buried in a family tomb. It is *possible* that Mary Magdalene was a family member. It is even *possible* that she was married to Jesus. But we can have no certainty that she was. Most historians would estimate a probability *substantially* less than 1 for these possibilities. Feuerverger's analysis assumes that Mary Magdalene should be in the tomb and his computation achieves statistical significance *only if she is assumed to be in the tomb*. This introduces another very serious source of statistical bias into his computations.

**4. A calculation using Bayes' theorem.** It is beyond the scope of this short comment to give full details on a more correct calculation. This journal has given us space on its web site for a 29 page article that defines the statistical issues of the tomb and then describes a series of calculations we have performed. Here, we will merely summarize the results of that article [Ingermanson (2008)].

We define the two events $J$ and $T$ as follows:

$J$ = the "Jesus son of Joseph" in the Talpiot tomb refers to Jesus of Nazareth,

$T$ = the observation of the rest of the Talpiot tomb data.

We denote the negation of the event $J$ by the symbol $\sim J$.

We are interested in computing the conditional probability $P(J|T)$ using Bayes' theorem:

$$P(J|T) = \frac{P(T|J)P(J)}{P(T|J)P(J) + P(T|\sim J)P(\sim J)}.$$

Define the two ratios

$$\alpha \equiv \frac{P(\sim J)}{P(J)},$$

$$\beta \equiv \frac{P(T|\sim J)}{P(T|J)}.$$

Then our formula simplifies to

$$P(J|T) = \frac{1}{1 + \alpha\beta}.$$

The results of many computations can be summarized as follows: $\alpha$ tends to be large, while $\beta$ is near 1. Therefore, $P(J|T)$ tends to be small.

We can estimate $\alpha$ very quickly. Feuerverger quotes the results of Camil Fuchs (2004) that the number of adult males who died in Jerusalem in the relevant time period was about 36420. This is overly precise, but it is reasonable in magnitude.

Assuming that 4% of men were named Jesus and 8.8% were named Joseph, we estimate the number of men named "Jesus son of Joseph" to be about 128. One of these men was Jesus of Nazareth. The other 127 are unknown to history.

Therefore, if we are given a randomly chosen man of Jerusalem named "Jesus son of Joseph," the probability that he is Jesus of Nazareth is $P(J) = 1/128$. The probability that he is not is $P(\sim J) = 127/128$. Taking the ratio, we estimate $\alpha \approx 127$. In general, if there were $N_J$ men of Jerusalem named "Jesus son of Joseph," then we have $\alpha = N_J - 1$.

The estimation of $\beta$ is much more complicated and we describe it in detail in the supplemental article [Ingermanson (2008)]. The general procedure is as follows:

We are comparing two hypotheses, $J$ and $\sim J$, using the data $T$ to distinguish between the two. For each of these two hypotheses, we imagine a statistical ensemble of tombs "similar" to the Talpiot tomb. We'll make random draws from each ensemble and tabulate the frequency of "hits" (random draws that agree with the data $T$).

We'll stipulate that each member of these two ensembles should contain an ossuary inscribed with "Jesus son of Joseph" and a second ossuary inscribed with "Judah son of Jesus." It should also contain two ossuaries bearing female names, two ossuaries bearing male names, and four uninscribed ossuaries. The distribution of names on the inscribed ossuaries must match the distribution of the names of persons living in Jerusalem in the first century, subject to the constraints of the two hypotheses.

In the case of the $\sim J$ hypothesis, there are no constraints.

In the case of the $J$ hypothesis, the only constraint is that the tomb must contain at least the names of certain members of the family of Jesus, with any remaining slots in the tomb filled with names chosen using the distribution of names in Jerusalem.

The procedure outlined above is similar in spirit to that followed by Feuerverger. Here are the primary differences in our calculations. We say that:

(1) The name of the mother of Jesus could have been inscribed as *any* form of Mary, including "Marya," "Mariam," or any other variant (even including the much-debated "Mariamenou Mara" inscription).

(2) "Judah son of Jesus" is considered less likely to appear in the tomb of Jesus of Nazareth than in the tomb of a randomly selected "Jesus son of Joseph."

(3) Jesus of Nazareth is considered less likely to be buried in a rock-cut tomb than was a randomly selected "Jesus son of Joseph."

(4) Mary Magdalene is not assumed to be in the tomb, and the "Mariamenou Mara" inscription is not assumed to be an appellation that applies to her with any higher probability than to any other Mary of Jerusalem.

(5) The probability of finding a Yoseh in the tomb is reckoned to be higher than usual, because the patriarch of the Talpiot family is named Joseph.

(6) The measure of "surprisingness" is the count of family members in the tomb, not Feuerverger's RR values. We use six different ways of defining this count.

The calculation was performed in Java using a wide variety of assumptions for the composition of a "Jesus family tomb" and using six different definitions of "surprisingness." Random draws were made in groups of 10,000, and results were tabulated.

The baseline calculation returned an estimate for the upper bound of $P(J|T)$ at about 2% (with a standard deviation of about 2%). A number of variants were tried, and the highest value found for $P(J|T)$ was 5.67%, using one assumption we consider unlikely. (The assumption that Yoseh should be exactly as rare in the Talpiot tomb as it is in tombs that do not have a patriarch named Joseph.)

We found that by tightening two assumptions, the upper bound could be substantially reduced. These are as follows.

We have assumed that the relative probability $\rho_{\text{son}}$ that Jesus had a son (as compared to other men of his time) was less than 1. That is, we defined a random variable $\rho_{\text{son}}$ uniformly distributed on the interval $[0,1]$. Many historians would argue that this distribution should be strongly weighted toward zero. Doing so would strongly reduce our estimates of $P(J|T)$.

Likewise, we have assumed that the relative probability $\rho_{\text{tomb}}$ that Jesus was reburied in a rock-cut tomb (as compared to other men of his time) was also less than 1. We defined a random variable $\rho_{\text{tomb}}$ uniformly distributed on the interval $[0,1]$. As noted earlier, Jodi Magness (2007) has argued strongly that $\rho_{\text{tomb}}$ should be heavily weighted toward zero. Doing so would again sharply reduce our estimates for $P(J|T)$.

We leave it to historians and archaeologists to debate such matters. We expect that their conclusions will tend to reduce our upper bound for $P(J|T)$ to be less than 2%, but it is impossible to predict how far it might drop. Such matters are irreducibly subjective.

DISCUSSION 7**5. Conclusion.** Feuerverger's computation contains a number of statistical biases, each of which favors $H_1$. One of these (the "Mariamenou" inscription) introduces an illicit factor of $1/44$ to RR, which accounts for a very strong bias all by itself. But five other factors enter in with moderate statistical bias toward $H_1$, and the net effect is to create the appearance of statistical significance where none actually exists.

We have performed a series of calculations using Bayes' theorem that estimate a likely upper bound for the probability that the Talpiot tomb is the tomb of Jesus of Nazareth. This upper bound is about 2% with a standard deviation of about 2%.

**Acknowledgments.** I thank Jay Cost for reading this article and making comments. I also thank Richard Bauckham, Mark Goodacre, Gary Habermas, Michael Heiser, Stephen Pfann, and James Tabor for helpful discussions over the last several months. I take full credit for any errors.## SUPPLEMENTARY MATERIAL

**Analysis of the Talpiot tomb using Bayes' Theorem and random variables** (doi: 10.1214/08-AOAS99GSUPP; .pdf). We analyze the Talpiot tomb, which has been alleged to be the family tomb of Jesus of Nazareth. Using Bayes' Theorem, we derive a simple function that estimates the probability that the tomb houses the remains of Jesus and his family. Unfortunately, this function cannot be evaluated exactly, because several of the key parameters are unknown. By using random variables with reasonable probability distributions, we examine the mean behavior and range of the function under a variety of conditions. We conclude that the probability is low (on the order of 2% or less) that the Talpiot tomb is the family tomb of Jesus of Nazareth.

## REFERENCES

BAUCKHAM, R. (2007). Guest article on Chris Tilling's blog on March 1, 2007. Available at http://www.christilling.de/blog/2007/03/guest-post-by-richard-bauckham.html.

BOVON, F. (2007). Article on the Society of Biblical Literature web site, posted March, 2007. Available at http://sbl-site.org/Article.aspx?ArticleId=656.

FUCHS, C. (2004). Democracy, literacy and names distribution in ancient Jerusalem—how many James/Jacob son of Joseph, brother of Jesus were there? *Polish J. Biblical Research* 1–30.

INGERMANSON, R. (2008). Supplement to "Discussion of: Statistical analysis of an archeological find." DOI: 10.1214/08-AOAS99GSUPP.

JACOBOVICI, S. and PELLEGRINO, C. (2007). *The Jesus Family Tomb*. HarperOne, San-Francisco.

MAGNESS, J. (2007). Article on the Society of Biblical Literature web site, posted March, 2007. Available at http://sbl-site.org/Article.aspx?ArticleId=640.

PFANN, S. (2007). Available at http://www.uhl.ac/Lost_Tomb/HowDoYouSolveMaria/.

2210 W. Main Street  
Suite 107, Box 103  
Battle Ground, Washington 98604  
USA  
E-mail: randy@rsingermanson.com  
URL: www.Ingermanson.com